# Electrosynthetic control of CNT conductivity & morphology: Scale-up of the transformation of the greenhouse gas $CO_2$ into carbon nanotubes by molten carbonate electrolysis

Stuart Licht[1,*], Matthew Lefler[1] Jiawen Ren[1], and Juan Vicini[1]

[1]Department of Chemistry, George Washington University, Washington, DC, 20052, USA. *mail: slicht@gwu.edu

**Abstract**

Transformation of carbon dioxide into carbon nanotubes, CNTs, by electrolysis in molten carbonates provides a low cost route to extract and store this greenhouse gas. CNTs are more stable, compact and valuable than fuels or other $CO_2$ conversion products, providing an incentive to remove $CO_2$ for climate mitigation. Previously, solid core carbon nanofibers, CNFs were formed with C-13 isotope $CO_2$, whereas hollow core fibers - carbon nanotubes, CNTs, were formed with natural isotope $CO_2$ splitting in molten lithium carbonate. Here we demonstrate the extraordinary range of specific morphologies and conductivities of CNTs that can be achieved through control of the electrolysis conditions in a one pot-synthesis, and scale-up of this process by which the greenhouse gas $CO_2$ is transformed into carbon nanotubes by molten carbonate electrolysis. Addition of $Li_2BO_3$, boron dopes and greatly enhances CNT conductivity formed at the galvanized steel cathode. Addition of $CaCO_3$ to the $Li_2CO_3$ electrolyte, decreases oxide solubility in the region of CNT growth, producing straight thin-walled CNTs. As we've previously shown, excess electrolytic oxide yields tangled CNTs, while addition of up to 50 mol% $Na_2CO_3$ to the $Li_2CO_3$ electrolyte, decreases electrolyte costs and improves conditions for intercalation in Na-ion anodes., and addition of $BaCO_3$ greatly increases electrolyte density opening pathways to the (free floating) extraction of electrogenerated CNTs, and finally that longer electrolysis time leads to proportionally wider diameter CNTs.

# Electrosynthetic control of CNT conductivity & morphology: Scale-up of the transformation of the greenhouse gas $CO_2$ into carbon nanotubes by molten carbonate electrolysis

Stuart Licht[1,*], Matthew Lefler[1] Jiawen Ren[1], and Juan Vicini[1]

[1]Department of Chemistry, George Washington University, Washington, DC, 20052, USA. *mail: slicht@gwu.edu

Carbon nanofibers, CNFs, or carbon nanotubes, CNTs, which are CNFs with hollow core, due to their superior strength, electrical and thermal conductivity, flexibility and durability have great potential as a material resource, with applications ranging from reinforced composites[1,2], capacitors[3], Li-ion batteries[4], nanoelectronics[5], and catalysts[6] to the principal component of lightweight, high strength building materials. Organo-metallic reactants using chemical vapor deposition, or arc discharge, or polymer pulling/carbonization are among the principal worthwhile, but costly methods of production of carbon nanofibers, CNFs or CNTs[7-9]. However, CNFs or CNTs formed from $CO_2$ as the reactant, could contribute to lower greenhouse gases for example by consuming, rather than emitting $CO_2$[10], storing carbon in a compact, stable form, and by providing a stable, carbon composite material that can be used as an alternative to steel, aluminum, and cement whose productions are associated with massive $CO_2$ emissions[11-13]. Increased availability of carbon composites will decrease $CO_2$ emissions by facilitating, both wind turbines and lightweight, low-carbon-footprint transportation[14]. Today CNT demand is mainly limited by the complexity and cost of the synthetic process, which requires 30 to 100 fold higher production energy compared to aluminum[15-17].

In 2009, we provided a theoretical derivation that $CO_2$ can be split by electrolysis at high solar efficiency[18], and demonstrated this high efficiency in 2010 through solar electrolysis of $CO_2$ in molten alkali eutectic carbonate or lithium carbonate to either solid (amorphous, or graphic) carbon at T < ~850°C and to carbon monoxide, CO, at T > ~850°C[19]. In the lower temperature domain, $CO_2$ removal occurs by the electrolytic reduction of tetravalent carbon in proximity to the cathode to solid, zerovalent carbon. The concentration of tetravalent carbon available for reduction is much higher as the electrolytic carbonate species, than as airborne or soluble $CO_2$. Molten carbonate, such as $Li_2CO_3$(liquid), contains ~10 molar reducible tetravalent carbon / liter. Air contains 0.04% $CO_2$, equivalent to only $1 \times 10^{-5}$ molar of tetravalent carbon / liter. Hence, molten carbonates formed by the dissolution of atmospheric $CO_2$, and its conversion to carbonate, provide a spontaneous, million-fold concentration increase of reducible tetravalent carbon sites per unit volume, compared to air, facilitating the direct conversion of atmospheric $CO_2$.

Until recently, CNTs had not been produced at low energy, nor had they produced in high yield from $CO_2$. In 2015 we introduced the efficient transformation splitting of $CO_2$, by electrolysis in molten carbonates to form carbon nanotubes and nanofibers at high yield. The electrolysis occurs at low electrical energy and high coulombic efficiency (~4 Faraday per mole $CO_2$)[10].

In 2015 we demonstrated that specific transition metals at the cathode, such as Ni, Co, Cu and Fe, act as nucleation points for high yield carbon nanotube growth[4,10,17]. The addition of Zn, such as occurs in a coating on galvanized steel, further lowers the activation energy for CNT formation[10,20]. In this process, the production of carbon, such as CNTs, by electrolysis in lithium carbonate occurs simultaneously with the production of oxygen and dissolved lithium oxide:

$$Li_2CO_3(liquid) \rightarrow C(CNT) + Li_2O(dissolved) + O_2(gas) \qquad (1)$$

$Li_2CO_3$ consumed by electrolysis is continuously replenished by reaction of this excess $Li_2O$, formed as a product in the Equation 1 electrolysis, with $CO_2$ from the air (or $CO_2$ available in higher concentration from stack emissions):

$$Li_2O(dissolved) + CO_2(gas) \rightarrow Li_2CO_3(liquid) \qquad (2)$$

for the net reaction (combining Equations 1 and 2):

$$CO_2(gas) \rightarrow C(CNT) + 1/2 O_2(gas) \qquad (3)$$

As illustrated in Figure S1 (Supporting Information), the synthesis splits atmospheric or smokestack $CO_2$, dissolved in a molten carbonate, by electrolysis. During the electrolysis, oxygen is evolved at a nickel anode, and with control of low levels of transition metal additives, uniform carbon nanotubes are produced at high yield at a steel cathode. Carbon isotope labeling was used to track $CO_2$ as the building blocks of the CNFs[17]. Solid core CNFs were formed with C-13 isotope $CO_2$[17], whereas more valuable carbon nanotube, CNT, morphologies were formed with inexpensive, natural isotope $CO_2$. This electrolytic synthesis provides a facile route for the transformation of theisgreenhouse gas into a wide range of high-value CNT commodities, and this wide range of valuable products formed at low energy to provide an important economic incentive to drive climate mitigation by carbon dioxide removal. The utility of CNTs will vary widely with the physical properties (morphology, uniformity, conductivity, etc.).

Here we expand the large range of specific morphologies, as summarized in Figure 1 and conductivities of CNTs that can be achieved through control of the electrolysis conditions in a one-pot molten carbonate synthesis. The top row of the figure presents SEM of carbon nanotubes synthesized from molten lithium carbonate. $CO_2$ is electrolytically split at high yield to oxygen gas at the anode, and carbon at the cathode. In the middle panel of the top row of Figure 1, Ni metal nucleation sites are evident as bright spots at the CNT terminals. In the right side SEM, CNTs prepared by extended duration electrolysis (ten hours, rather than 1 hour) are substantially thicker diameter. In the middle row of the figure on the left side, electrolysis in C-13 isotope carbon salts ($^{13}CO_2$ in $Li_2{}^{13}CO_3$) form more solid core carbon nanofibers, rather than carbon nanotubes, and on the right (panel D), upon dissolution of lithium oxide, $Li_2O$, into (natural abundance isotope) $Li_2CO_3$, forms by electrolysis tangled, rather than straight, CNTs are formed. The tangled CNTs result from $sp^3$ defects in the $sp^2$ graphene layered structures, as detected by Raman spectroscopy, and can lead to improved rechargeable battery storage capacity[4]. The bottom of the figure shows the SEM of CNTs grown at high yield by electrolysis in molten mixed carbonates, the left utilizes a less expensive electrolyte (50/50 wt% of $Na_2CO_3$ mixed with $Li_2CO_3$), and the right side uses a more dense electrolyte (20/80 wt% of $BaCO_3$ mixed with $Li_2CO_3$).

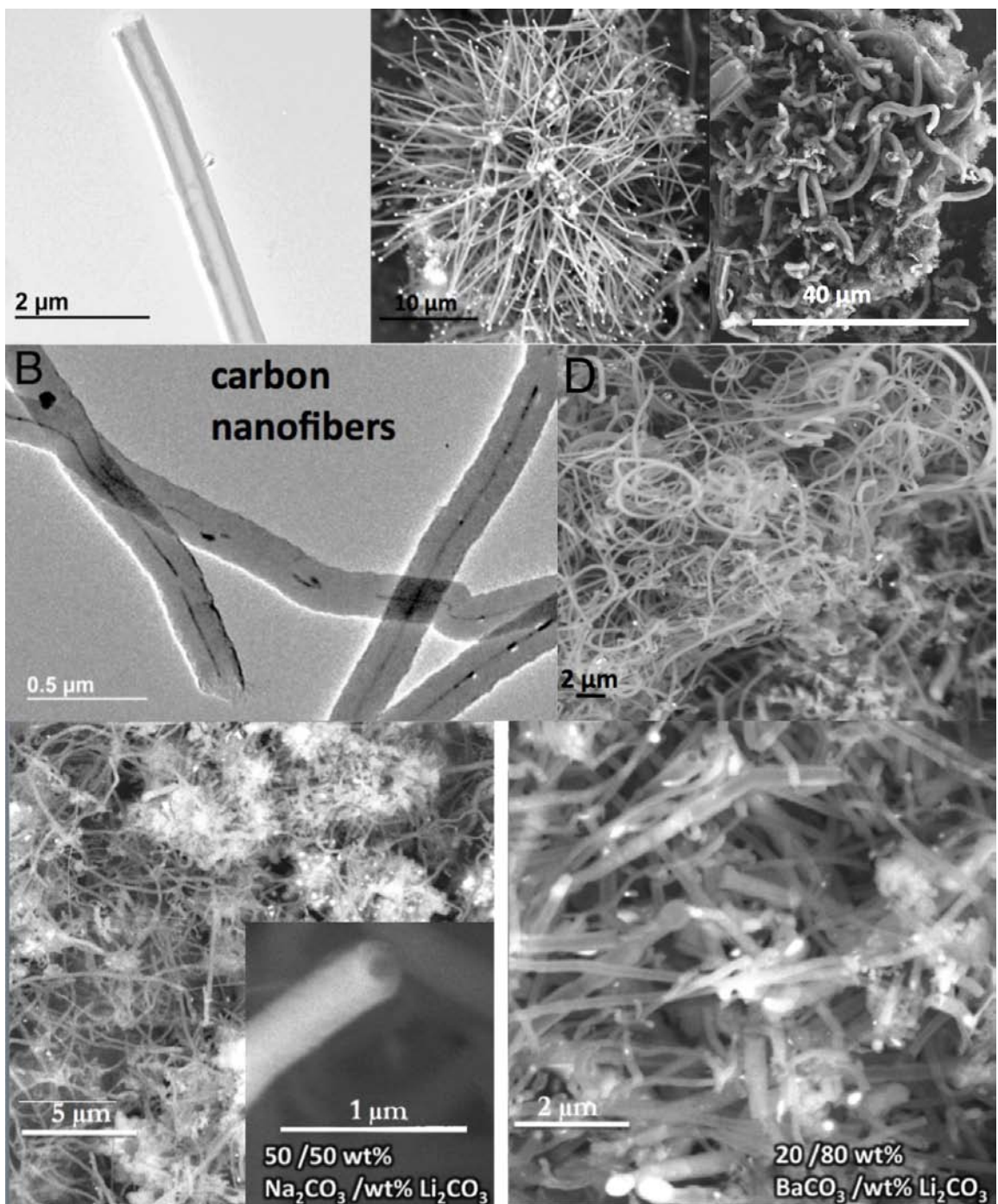

Figure 1 Portfolio of CNT and CNF morphologies synthesized at high yield from carbon dioxide in molten carbonate under various electrolysis conditions. Top row: carbon nanotubes are electrosynthesized molten lithium carbonate during short (middle) or long term (right) electrolyses. The middle row, left: solid core carbon nanofibers are grown from $^{13}$C isotope, while right: tangled, rather than straight carbon nanotubes are grown from lithium carbonate containing dissolved oxide. Bottom: A high yield of carbon nanotubes can also be grown by electrolysis from mixed Na/L (left) or Ba/LI carbon electrolyses[10,17,28].

**RESULTS AND DISCUSSION**

**One-pot synthesis of boron-doped highly conductive CNTs**

In 1999, Wei and co-workers demonstrated that boron doping of bulk arc synthesized CNTs can improve their room temperature electrical conductivity by an order of magnitude[21], which has been subsequently confirmed for both arc and CVD synthesized carbon nanotubes[9,22-25].

Pure $B_2O_3$ (mp 450 °C, white, melts clear), but is a glass insulator, while dissolved $Li_2O$ in $B_2O_3$, as $LiBO_2$, is an electrochemical conductive liquid, which we have previously explored as a molten air battery electrolyte[26]. The binary system of $B_2O_3$ and $Li_2O$ (mp 1438 °C, white, dissolves clear) presents a complex phase diagram with an extensive homogenous liquid phase above 767 °C and is included as Supporting Information Figure S2. We observe here that the combined salt lithium metaborate, $LiBO_2$ (mp 849 °C, white) is highly soluble in $Li_2CO_3$ (dissolves clear), and is a successful additive for the one-pot synthesis of boron-doped highly conductive CNTs. We measured that our tangled synthesized with $Li_2CO_3$ and $Li_2O$, and known to have defects[4,10], have a conductivity comparable to purchased industrial grade carbon nanotubes, and our straight CNTs synthesized with pure $Li_2CO_3$ have a conductivity three-fold higher. Relative to this our boron-doped CNTs synthesized here with a 10% wt of $LiBO_2$ in the electrolyte have a measured conductivity 6-fold higher than our straight CNTs synthesized with only pure $Li_2CO_3$, and no additives. Interestingly, 20 wt% $LiBO_2$ exhibits a lower conductivity (and measured as intermediate to that between the straight and tangled CNTs). These 20 wt% synthesized samples also exhibit a high level concentration of amorphous carbon suggesting that the maximum CNT conductivities will occur at less than 10% wt of $LiBO_2$. The discrepancy between boron-doping and higher than 10 wt % $LiBO_2$ in the electrolyte, appear to be consistent with competition between carbonate and metaborate reduction at the cathode, and that the latter can dominate at higher metaborate concentrations. The standard voltage of the reduction of lithium metaborate is ~2.0V according to thermodynamic data, which is marginally higher, but competitive with the reduction of carbonate to carbon.

$$2LiBO_2 \rightarrow 2B + Li_2O + 3/2O_2; \qquad E=-G/nF = 2.0V \text{ at } 1100K \qquad (4)$$

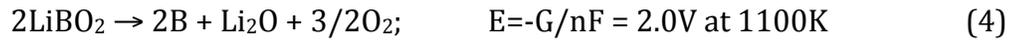

The top of Figure 2 shows typical SEM images for B-doped CNTs synthesized from a 5g $LiBO_2$ plus 50g $Li_2CO_3$ (10 wt%) electrolyte. Despite some amorphous carbon nanoparticles, a majority of CNTs are evident in the product. Compared to our previous syntheses of CNF/CNTs using pure lithium carbonate, there is a higher content of amorphous carbon particles, indicating that the involvement of lithium metaborate makes the syntheses less homogeneous, possibly due the competing effect of B-deposition with nickel nucleation site. When higher concentrations than 10 wt% of lithium metaborate are present, for example, from an electrolyte composed of 10g of lithium borate with 50g (20 wt%) , the resulting product has a much lower content of carbon nanotubes as measured by SEM (Figure 2 bottom), evidencing the interference effect of too high a concentration of lithium metaborate towards CNTs synthesis.

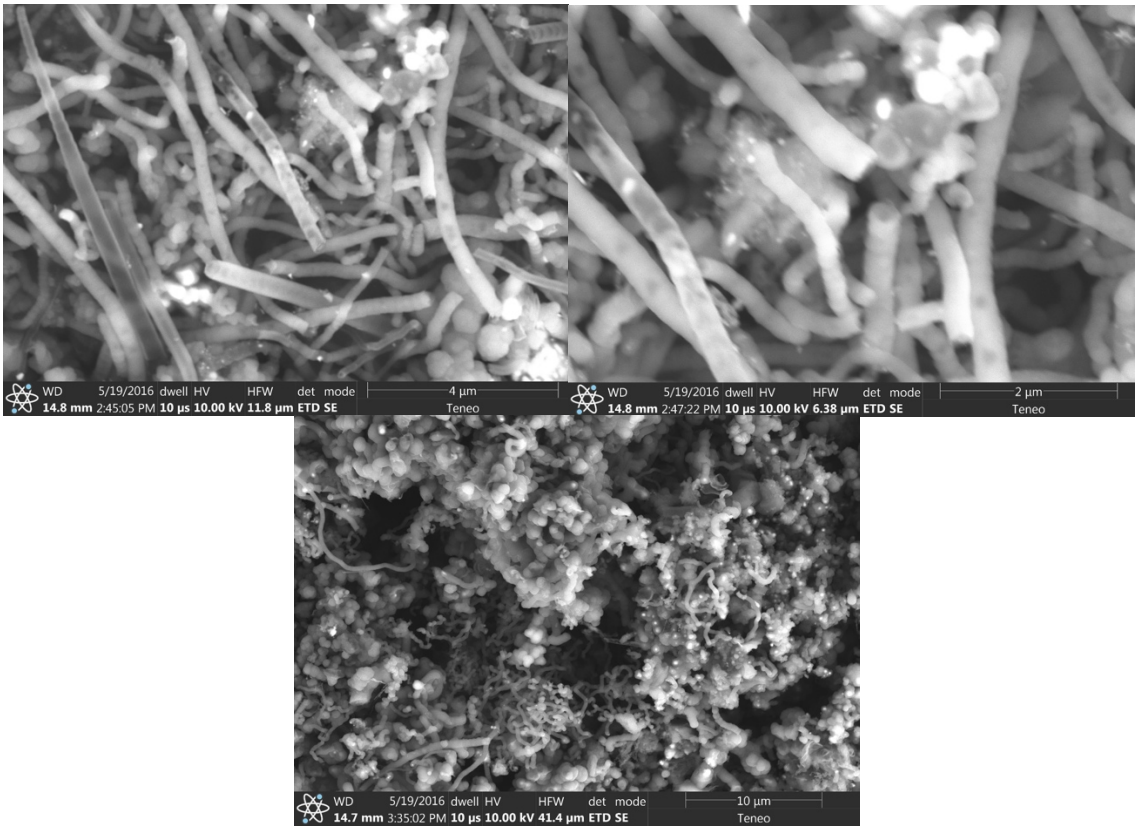

Figure 2. SEM of B-doped CNTs prepared with either a (top) 10 wt% or (bottom) 20 wt% $LiBO_2$ additive to the molten 770°C $Li_2CO_3$ electrolyte.

To verify that the obtained CNTs are boron-doped CNTs rather than a mixture of elemental boron plus pure CNTs, Raman spectra were recorded with an incident laser at 532 nm. The spectrum in Figure 3 exhibits a G-band shift to higher wavelength. CNTs fabricated from pure lithium carbonate electrolyte, exhibit an observed G band at 1575 cm$^{-1}$; however, for the 10 wt% $LiBO_2$ B-CNT product (from 5g of lithium metaborate + 50g of Lithium carbonate), the G band shits to 1591 cm$^{-1}$. The shift of G-band in this direction can reflect the content of doping of boron[32]. This spectrum indicates that the 10 wt% lithium metaborate molten carbonate synthesized sample is a boron-doped carbon rather than a mixture of elemental boron and pure CNTs.

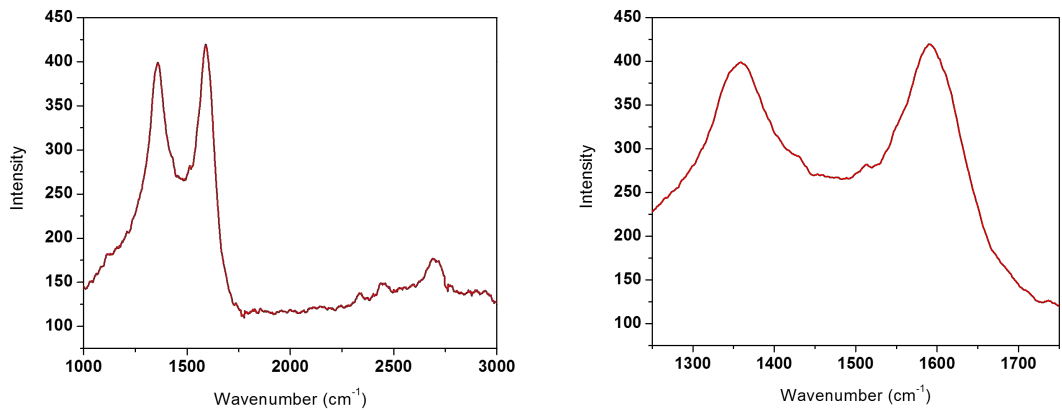

Figure 3. Raman spectrum of B-doped CNTs prepared with a 10 wt% LiBO$_2$ additive to the molten Li$_2$CO$_3$ electrolyte over an expanded (left) and restricted (right) frequency range.

**Calcium carbonate induced synthesis of thin walled CNTs**

Interestingly, we observed that while the aqueous solubility of calcium oxide is high, and while aqueous calcium carbonate is nearly insoluble, the opposite solubility trend occurs in molten lithium carbonate[27]. As shown in Figure S3 of the Supporting Information, the solubilities of either lithium oxide, Li$_2$O, or barium oxide, are high (> 6 molal) in Li$_2$CO$_3$ at 750°C, whereas CaO is nearly insoluble. Normally, CaCO$_3$ does not melt, decomposing to CaO and CO$_2$ at ~850°C in the conventional limestone to lime production process. However, at lower temperatures we found that CaCO$_3$ is highly soluble in molten Li$_2$CO$_3$. As also seen in that figure CaCO$_3$ is highly soluble (> 6 molal at 750°C) in molten Li$_2$CO$_3$, and can provide an enriched source of Ca$^{2+}$ cations prior to electrolysis. When these soluble Ca$^{2+}$ cations are available prior to electrolysis, CaO insolubility occurs during electrolysis:

$$\text{CaCO}_3(\text{liquid}) \rightarrow \text{C(CNT)} + \text{CaO(solid)} + \text{O}_2(\text{gas}) \qquad (5)$$

This CaO insolubility provides an opportunity to substantially modify the interfacial environment at the electrolyte/cathode interface during carbonate electrolysis. CNTs are grown in this unusual high CaO environment, and CaO can both intermingle with the CNTs during formation, and excess CaO can precipitate to the bottom of the electrolysis chamber (due to the higher density of CaO than the Li/Ca carbonate mix electrolyte).

As shown in Figure 4, it is interesting that thin-walled CNTs are synthesized when a sufficient amount of $CaCO_3$ is added to the electrolytes. A 10g of $CaCO_3$ plus 50g of $Li_2CO_3$ electrolyte provided a chemical environment for the production of thin-walled CNTs. Compared to Figure 1 top left, the wall thickness of CNTs synthesized in pure $Li_2CO_3$ is ~200nm, which is much thicker than CNT-Ca which is ~50nm.

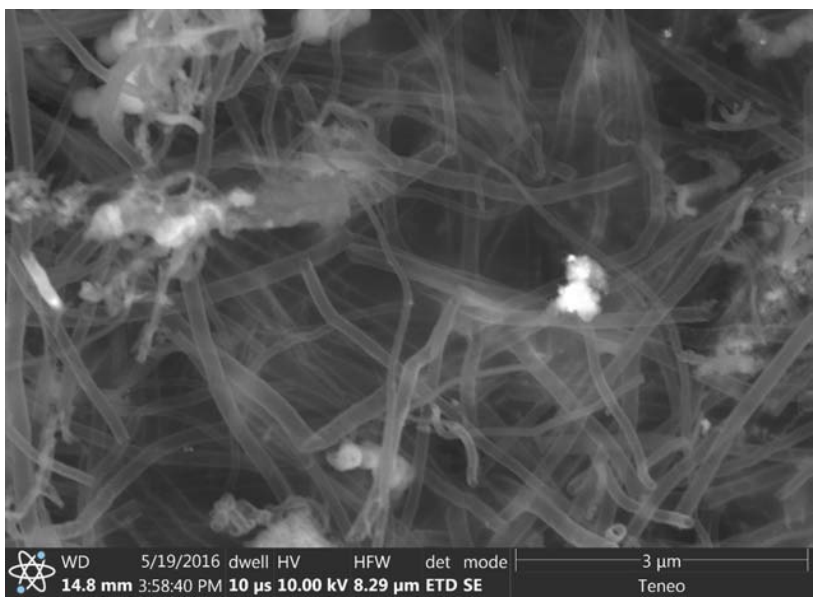

Figure 4. SEM of CNTGs prepared with 20 wt% $CaCO_3$ added to the molten 770°C $Li_2CO_3$ electrolyte.

**Scale-up of $CO_2$ transformation to CNTs by molten carbonate electrolysis**

One mode of $CO_2$ to carbon nanotube transformation by molten carbonate electrolysis utilizes hot $CO_2$ that would have been emitted from a fossil fuel power plant, and uses a portion of the plant electric power to convert all of the $CO_2$ into CNTs. Such a process simplifies the design by eliminating the need to heat the electrolysis or externally power (for example by renewable energy) the electrolysis chamber. One such design consists of the molten electrolyte circulated by pump through separate hot $CO_2$ addition and electrolysis chambers as shown in Figure 5. $CO_2$ dissolution occurs in accord with equation 2. $Li_2O$ is continuously formed and dissolves in the electrolyte in the electrolysis chamber in accord with equation 1.

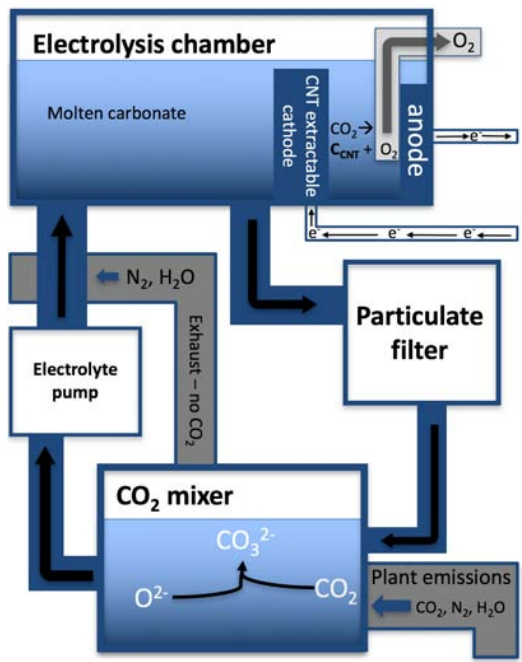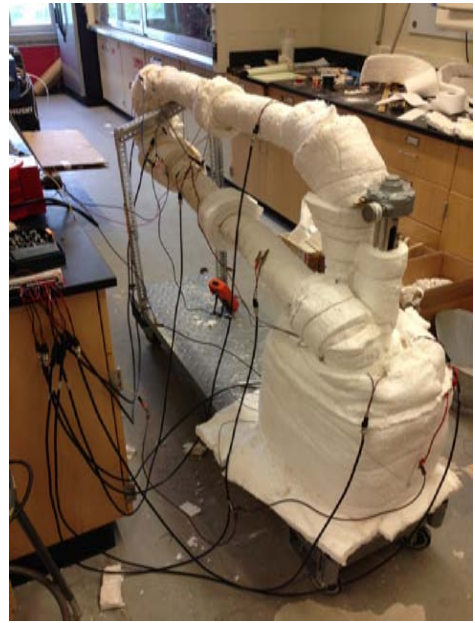

Figure 5. Left: $CO_2$ to CNT transformation using a pumped, circulated molten carbonate electrolyte, a $CO_2$ mixing chamber inputting hot flue gas, dissolving the CO2 and emitting the insoluble $N_2$ and $H_2O$ from a standard flue stack, and an electrolysis chamber producing carbon nanotubes at the cathode and releasing $O_2$ at the anode. Right: In-lab molten salt loop using a Wenesco HTPA2C molten salt pump.

The electrolysis chamber is being rapidly increasing in size and CNT output. The chamber is linearly scalable with electrode surface area. As seen in Figure 6, early molten carbonate electrolyses were conducted within a cylindrical alumina crucible and (shown on the mid-bottom of the photo) utilized coiled Ni wire anode above a coiled galvanized steel wire cathode. Each had a surface area of 5 cm² (10 cm² exposed above and below the coil), and were scaled up to 800 cm² electrodes in larger crucibles, as well as in a modified cell in which the crucible is nickel, rather than alumina, and the interior nickel walls served as both container and oxygen generation anode[10].

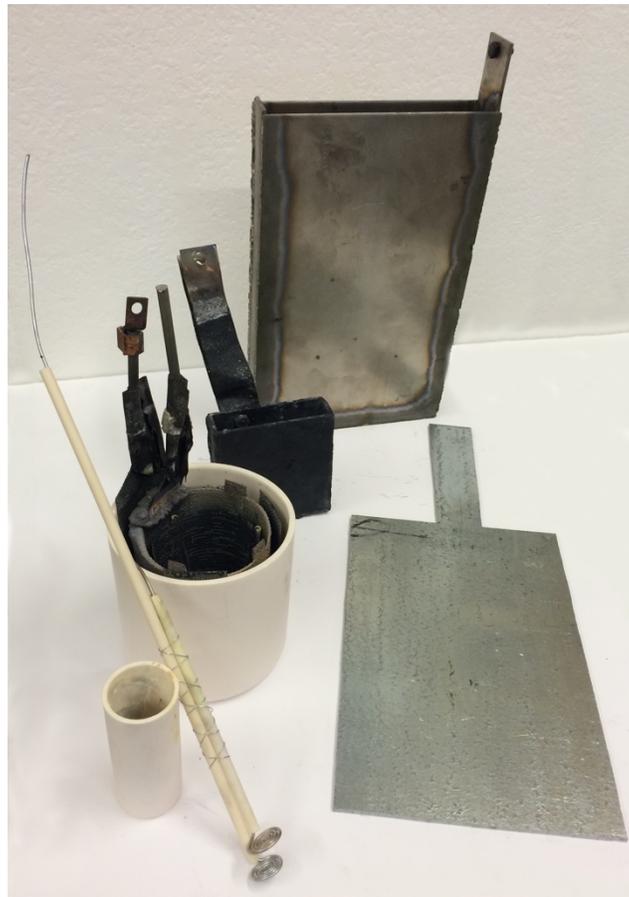

Figure 6. The evolution of the electrolysis chamber. Earlier versions can be seen in the front on the left, and later versions in the back and to the right. Size and morphology change as the current used in an electrolysis increases well beyond the capabilities of previous versions.

We accomplish a decrease in the required electrolyte volume, as shown in Figure 6, by moving from a cylindrical to a rectangular electrolysis chamber. Both nickel body (shown) and graphite body electrolysis chambers have been built. The graphite body exposed to the hot air needs to be protected, such as through external ceramic tiles or an alumina cement overlayer to prevent oxidative corrosion. The nickel body electrolysis cell is highly stable against corrosion and forms a stablizing black nickel oxide overlayer during the first heating and electrolysis. The smaller rectangular cell shown in the figure is shown following seven successive electrolyses, including an extended 10 hour electrolysis. The electrolyses in the nickel rectangular cells used the interior walls of the nickel as air electrodes, and subsequent to a change to black, and expansion by < 200 microns per side (due to the oxide formation) during the first electrolysis, the 3mm width of the nickel plate did not measurably change during subsequent electrolyses. Rather than previous electrolyses in which the cathode was cooled prior to product extraction (and removed either by uncoiling the cathode wire, in which case the product drops off, or by hammer hitting the cylindrical sheet cathode electrodes[10]), here the product is extracted while the electrode is hot by sliding another stiff steel sheet along the surface of the cathode. Combined with this process, centrifugal spinning, or denser electrolytes, such as those including barium carbonate[28], can yield alternative extractions of the CNT product.

The largest cell shown in Figure 6 is prior to its first electrolysis. The larger cell is wider to test the effect of the inclusion of multiple electrodes (in this case, up to two cathodes sandwiching a third (other than the walls) nickel anode in a single electrolysis chamber. Each of the multiple anodes or cathodes are shorted as an electrically contiguous single surface area of cathode or anode. A single one of these galvanized steel cathodes is shown in the right forefront of the figure. Not shown are groove-cut ceramic tube "bumpers" which prevent anode/cathode shorting. Two of these cathodes have an area of 2,000 cm² per cell, and with a 0.2 A cm$^{-2}$ are capable of the generation 1 kg of CNT per day while consuming 4 kg of $CO_2$ per day. In the electrolysis experiments, constant currents to ~100 A are provided by a Xantrex 8-100 power supply and higher currents by a Power Ten Inc. 660A power supply. In these step-wise scale-up experiments, 10 m² contiguous electrodes utilize Ni clad on copper anodes and steel clad on copper anodes which evenly distribute current throughout the larger electrodes, to be capable of generating 50 kg of CNT per day and consuming 200 kg of $CO_2$ per day.

Higher rates of CNT generation and $CO_2$ consumption increase linearly with higher current density but require higher electrolysis voltage. Figure 7 presents the measured electrolysis voltage as a function of current and current density for simple (planar, equal surface area) 480 cm² electrodes. Electrolysis voltages are considerably lower, when the electrodes are not planar, but are additionally textured either at a macroscopic, microscopic or nanoscopic level, and with texturing will decrease linearly the effective current density with the increasing effective surface area. For example, macroscopic grooves spaced by 1 mm and cut 3 mm deep more than triple the effective surface area of the electrode and at constant applied current diminish the effective current density by more than a factor of three.

As delineated in our recent studies whereas the metered voltage to drive the electrolysis transforming $CO_2$ to CNTs may be 0.9 to 1.1 V, the actual voltage is 0.57 V due to efficiency improvements from the synergistic coupling of the electrolyzer to a combined cycle power plant[29,30]. Specifically, these consist of (1) heat liberated during dissolution of $CO_2$ into the molten carbonate, which increases the electricity generated by the steam turbine cycle, (2) decreased heat losses due to decrease volume of power plant stack emissions, and (3) improved efficiency of the gas turbine cycle by mixing in pure oxygen, rather than air into the gas turbine. This pure oxygen is evolved "free" during the splitting of $CO_2$ to $O_2$ and CNTs. This $O_2$ can provide one of the two $O_2$'s needs per methane combusted ($CH_4 + 2O_2 \rightarrow CO_2 + 2H_2O$). The direct quantification of these is by feeding the oxygen and added heat back into the NG CC power plant and measuring the efficiency increase.

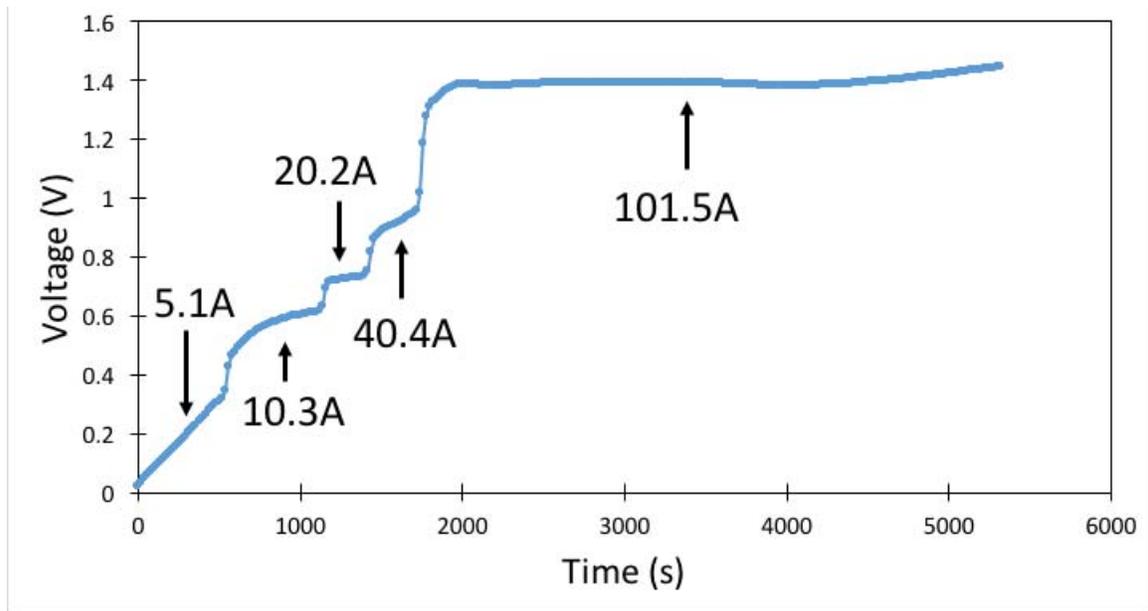

Figure 7. With a surface area of 480cm², the electrolysis was conducted in 770°C $Li_2CO_3$ at 5.1A (10.6mA/cm²) for 10 minutes, followed by 10.3A (21.5mA/cm²) for 10 minutes, then 20.2A (42.1mA/cm²) for 5 minutes, 40.4A (84.2mA/cm²) for 5 minutes, and finally a growth period of 1 hour at 101.5A (211.5mA/cm²).

**MATERIALS AND METHODS**

This study focuses on the wide variety of controlled molten carbonate electrolyzer conditions to form a carbon nanotubes with specific morphologies and properties and the scale-up of the carbon dioxide to carbon nanotube process. $^{13}C$ lithium carbonate (Sigma-Aldrich, 99 atom% $^{13}C$), $^{13}C$ carbon dioxide (Sigma-Aldrich, 99 atom% $^{13}C$), lithium carbonate (Alfa Aesar, 99%) and lithium oxide (Alfa Aesar, 99.5%) are combined to form various molten electrolytes. anhydrous lithium metaborate (LiBO$_2$, 99.9%, Alfa Aesar 12591), boron oxide (B$_2$O$_3$, 99.98%, Alfa Aesar 89964).

Electrolyses are driven galvanostatically, at a set constant current as described in the text. The electrolysis is contained in a pure nickel 100 ml crucible (Alfa Aesar). Electrolyses in the Ni crucible used the inner walls of the crucible as the anode. A wide variety of steel wires for coiled cathodes are effective; an economical form used in this study is a hardware store variety galvanized (zinc coated) steel wire, V2568 18 gauge (diameter 0.05 inch). Following an initial low current (0.05 A for 10minutes, 0.1A for 10 minutes, 0.2A for 5 minutes, and 0.4A for 5 minutes) step to grow Ni nucleation sites on the cathode, CNTs or CNFs are grown on an immersed 10 cm$^2$ galvanized steel cathode at a constant current of ~1 A for 1 hour. The electrolysis potentials are consistent with those we have recently reported for generic carbon deposition in a molten Li$_3$CO$_3$ electrolyte as a function of current density, and as previously reported, the potential decreases with addition of Li$_2$O to the electrolyte[3]. Two nanostructures are generated, straight CNTs or CNFs that are grown in electrolyte without added Li$_2$O, or tangled CNTs or CNFs that are grown

Li$_2$O has been added to the electrolyte. During electrolysis, the carbon product accumulates at the cathode, which is subsequently removed and cooled. Subsequent to electrolysis the product remains on the cathode, but falls off with electrolyte when the cathode is extracted, cooled, and uncoiled. The product is washed with 11 m HCl, and separated from the washing solution by centrifugation.

The carbon product is washed, and analyzed by PHENOM Pro-X Energy Dispersive Spectroscopy (EDS) on the PHENOM Pro-X SEM or FEI Teneo LV SEM and by TEM with a JEM 2100 LaB6 TEM. Raman spectroscopy was measured with a LabRAM HR800 Raman microscope (HORIBA) with 532.14 wavelength incident laser light, with a high resolution of 0.6 cm$^{-1}$.

In the syntheses, a constant mass of 50g lithium carbonate was utilized. For the Boron-doped CNF/CNTs, various amount of lithium metaborate was added, included 5g or 10g of LiBO$_2$ with 50g of Li$_2$CO$_3$, and those of 1.5g, 3g, and 8g, will be reported soon. To investigate cations influence on the morphology of the CNF/CNTs, CaCO$_3$ was added into the electrolyte. After melting at 770°C, the cathode was immersed into the electrolyte and the polarization pre-treatment was conducted (0.05A/10min, 0.1A/ 10min, 0.2A/5min, and 0.4A/5min) with subsequent electrolysis at a constant current of ~1A.

**CONCLUSION**

CNTs formed from CO$_2$ as the reactant, can contribute to lower greenhouse gases, for example by consuming, rather than emitting CO$_2$, storing carbon in a compact, stable form, and by providing a stable, valuable, carbon composite material. Despite their superior mechanical, thermal and electronic properties, carbon nanotubes applications had been limited to date by the high cost of their conventional chemical vapour deposition synthesis. This CNT cost is higher than for carbon fibers, and 2 to 3 orders of magnitude higher than that of conventional graphitic or amorphous carbon production[29]. Here, the facile, inexpensive synthesis of CNTs at high yield into a wide variety of uniform carbpon nanotubes by electrolysis in molten carbonate at cost effective nickel and steel electrodes, and provides a significant ecomomic impetus to transform the greenhouse gas carbon dioxide into a useful, stable, valuable commodity.

**References**

1. Sattar, R., Kausar, A., Siddiq, M. Advances in thermoplastic polyurethane composites reinforced with carbon nanotubes and carbon nanofibers: A review. *J. Plast. Film & Sheeting* **31**, 186-334 (2015).

2. Estili, M., Sakka, Y. Recent advances in understanding the reinforcing ability and mechanism of carbon nanotubes in ceramic matrix composites. Sci. Tech. Adv. Mat. **15**, 1-25 (2014).

3. Srikanth, V., Ramana, G. V., Kumar, P. S. Perspectives on State-of-the-Art Carbon Nanotube/Polyaniline and Graphene/Polyaniline Composites for Hybrid Supercapacitor Electrodes. *J. Nanosci. Nanotech.* **16**, 2418-2424 (2016).

4. Licht S., Douglas, A., Ren, J., Carter, R., Lefler, M., Pint, C. L. Carbon Nanotubes Produced from Ambient Carbon Dioxide for Environmentally Sustainable Lithium-Ion and Sodium-Ion Battery Anodes**.** ACS Central Sci. **2**, 162-168 (2016).

5. Che, Y. C., Chen, H. T., Gui, H., Liu, J., Liu, B. L., Zhou, C. W. Review of carbon nanotube nanoelectronics and macroelectronics. *Semiconductor Sci. Tech.* **29**, 073001 (2014).

6. Yan, Y., Miao, J., Yang, Z., Xiao, F.-X., Yang, H, B., Liu, B., Yang, Y. Carbon nanotube catalysts: recent advances in synthesis, characterization and applications**.** Chem.

7. Shah, K. A., Tali, B. A. Synthesis of carbon nanotubes by catalytic chemical vapour deposition: A review on carbon sources, catalysts and substrates. *Mat. Sci. Semiconductor Processing* **41**, 67-82 (2016).

8. Lichao, F., Xie, N., Zhong, J. Carbon nanofibers and their composites: A review of synthesizing, properties and applications. Mat. **7**, 3919-3945 (2014).

9. Mukhopadhyay, I., Hoshino, N. Kawasaki, F. Hsu, W. K., Touhara , H. Electrochemical Li Insertion in B-doped multiwall carbon nanotubes. *J. Electrochem. Soc.* **149**, A39-A44 (2002).

10. Ren, J., Li, F.-F., Lau, J., Ganzalez-Urbina, L. & Licht, S. One-pot synthesis of carbon nanofibers from $CO_2$. *Nano Lett.* **15**, 6142-6148 (2015).

11. Licht, S. Efficient Solar-Driven Synthesis, Carbon Capture, and Desalinization, STEP: Solar Thermal Electrochemical Production of Fuels, Metals, Bleach. *Adv. Mat.* **47**, 5592-5612 (2011).

12. Licht, S., Wu, H. STEP iron, a chemistry of iron formation without $CO_2$ emission: Molten carbonate solubility and electrochemistry of iron ore impurities. *J. Phys. Chem. C* **115**, 25138-25157 (2011).

13. Licht, S., Wu, H., Hettige, C., Wang, B., Lau, J., Asercion, J., Stuart, J. STEP Cement: S̲olar T̲hermal E̲lectrochemical P̲roduction of CaO without $CO_2$ emission. *Chem. Comm.* **48**, 6019-6021 (2012).

14. Song, Y., Youn, J., Gutowski, T. Life cycle energy analysis of fiber-reinforced composites. *Composites: Part A*, **40**, 1257–1265 (2009).

15. pluscomposites, *Composites: Materials of the Future:* - Part 4: Carbon fibre reinforced composites, at: http://www.pluscomposites.eu/publications, directly accessed July, 16, 2015 at:

http://www.pluscomposites.eu/sites/default/files/Technical%20series%20-%20Part%204%20-%20Carbon%20fibre%20reinforced%20composites_0.pdf

16. Kim, H. C.; Fthenakis, V. Life Cycle Energy and Climate Change Implications of Nanotechnologies. *J. Industr. Ecology*, **17**, 528–541 (2012).

17. Ren, J., Licht, S. Tracking airborne $CO_2$ mitigation and low cost transformation into valuable carbon nanotubes. *Sci. Rep.* **6**, 27760 ( (2016).

18. Licht, S. STEP (solar thermal electrochemical photo) generation of energetic molecules: A solar chemical process to end anthropogenic global warming. *J. Phys. Chem. C* **113**, 16283-16292 (2009).

19. Licht, S., et al. A New Solar Carbon Capture Process: STEP Carbon Capture. *J. Phys. Chem. Lett.* 1, 2363-2368 (2010).

20. Dey, G., Ren, J., El-Ghazawhi, T., Licht, S. How does amalgamated Ni cathode affect Carbon Nanotube growth? A density functional theory study. RSC Adv. **6**, 2016; 27191-27196 (2016).

21. Wei, B., Spolenak, R., Kohler-Redlich, P., Ruhle, M., Arzt, E. Electrical Transport in pure and boron doped carbon nanotubes. *Appl. Phys. Lett.* **74**, 3149-3151 (1999).

22. Charlier, J.-C., Terrones, M., Baxendale, M., Meunier, V., Zacharia, T., Rupesinghe, N. L., Hsu, W. K., Grobert, N., Terrones, H., Amaratunga, G. A. J. Enhanced Electron Field Emission in B-doped Carbon Nanotubes. *Nano Lett.* **2**, 1191-1195 (2002).

23. Lee, J. M., Park , J. S. Lee, S. W., Kim, H., Yoo , S., Kim. S. O., Selective Electron- or Hole-Transport Enhancement in Bulk-Heterojunction Organic Solar Cells with N- or B-Doped Carbon Nanotubes. *Adv. Mat.* **23**, 629-633 (2011).

24. Yang, L., Jiang, S., Zhao, Y., Zhu, L., Chen, S., Wang, X., Wu, Q., Ma, J., Ma, Y., Hu, Z. Boron-Doped Carbon Nanotubes as Metal-Free Electrocatalysts for the Oxygen Reduction Reaction. *Angew. Chem. Int..* **50**, 1-6 (2011).

25. Monteiro, F. H., Larrude, D. G. Maia da Costa, M. E. H., Terrazos, L. A., Capaz, R. B.

Freire, Jr., F. L. Production and characterization of boron doped single wall carbon nanotubes. *J. Phys. Chem. C* **116**, 3281-3285 (2012).

26. Licht, S., Cui, B., Stuart, J., Wang, B., Lau, J. Molten air – a new, highest energy class of rechargeable batteries. *Energy & Env. Sci.* **6**, 3646-3657 (2013).

27. Licht, S., Cui, B., Wang, B. STEP carbon capture – the barium advantage. *J. CO$_2$ Utilization*, **2**, 58-63 (2013).

28. Wu, H., Li, Z., Ji, D., Liu, Y., Yuen, D., Zang, Z., Ren, J., Lefler, M., Want, B., Licht, S. One-pot synthesis of nanostructured carbon materials from carbon dioxide via electrolysis in molte carbonate salts, *Carbon* **196**, 208-217 ( (2016).

29. Lau, J., Dey, G., Licht, S. Thermodynamic assessment of CO$_2$ to carbon nanofiber transformation for carbon sequestration in a combined cycle gas or coal power plant. *Energy Conversion Management*, doi: 10.1016/j.enconman.2016.06,007 ( (2016).

30. Ren., J., Lau, J., Lefler, M., Licht, S. The minimum electrolytic energy needed to convert carbon dioxide to carbon by electrolysis in carbonate melts. *J. Phys. Chem. C* **115**, 25138-25157 (2011).

31. cheaptubes.com, Industrial Grade Carbon Nanotubes. Nov. 1, 2014 at: http://www.cheaptubes.com/product-category/industrial-grade-carbon-nanotubes/

32. Ishii, S., Watanabe, T., Ueda, S., Tsuda, S., Yamaguchi, T., Takano, T. Resistivity reduction of boron-doped multiwalled carbon nanotubes synthesized from a methanol solution containing boric acid. *Appl. Phys. Lett.* **92**, 202116 (2008)

# Supporting Information

**Electrosynthetic control of CNT conductivity & morphology: Scale-up of the transformation of the greenhouse gas $CO_2$ into carbon nanotubes by molten carbonate electrolysis**

Stuart Licht[1,*], Matthew Lefler[1] Jiawen Ren[1], and Juan Vicini[1]

[1]Department of Chemistry, George Washington University, Washington, DC, 20052, USA. *mail: slicht@gwu.edu

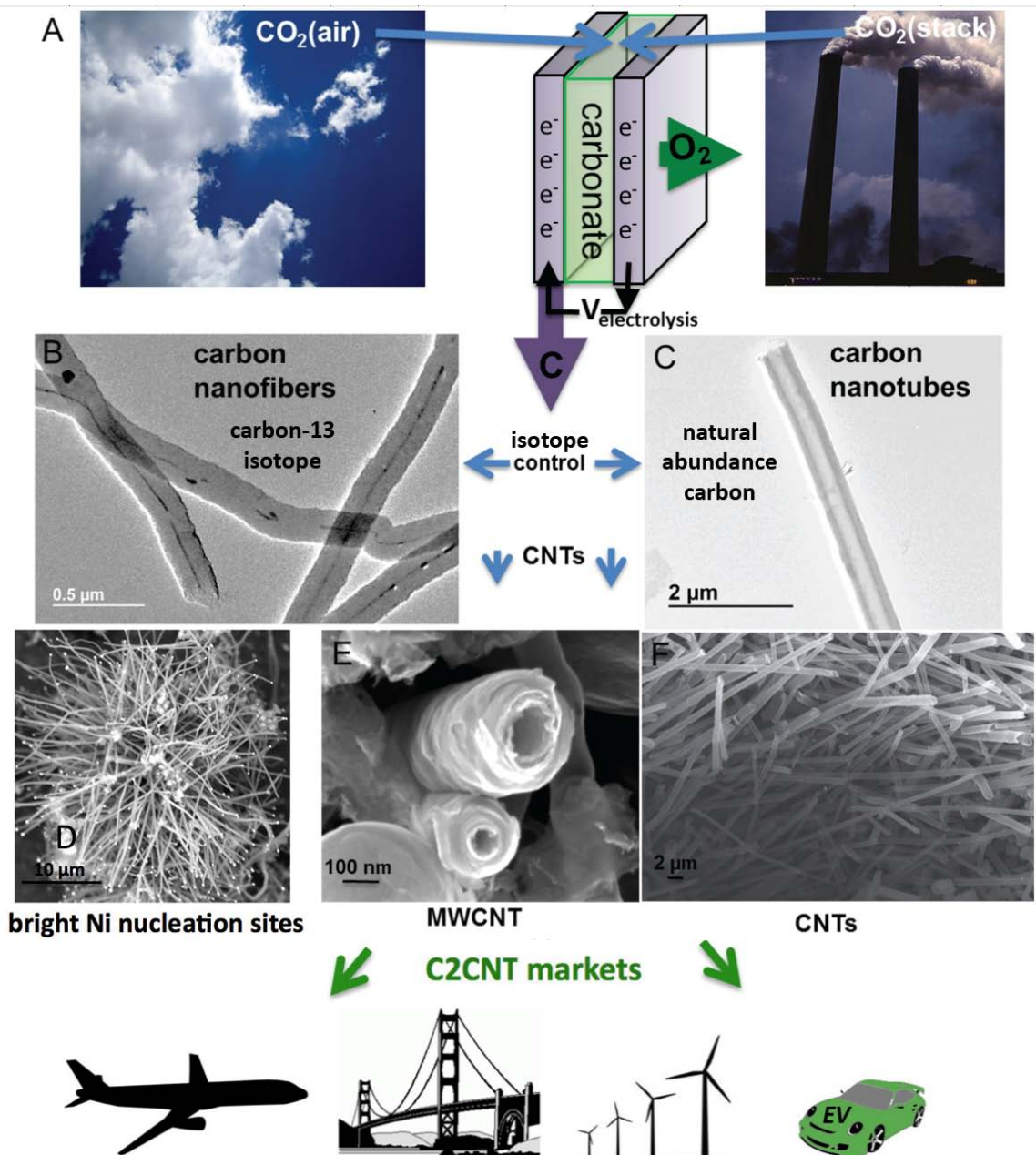

**Figure S1 (A)** High yield electrolytic synthesis of carbon nanostructures from dissolved air or smoke stack concentrations of $CO_2$ in molten lithium carbonate. (B,C) The carbon isotope controls formation of either nanofiber (B: as grown from $^{13}CO_2$ in $Li_2^{13}CO_3$) morphologies or nanotube morphologies (C: as grown from natural abundance $CO_2$ & $Li_2CO_3$). (D) During $CO_2$ electrolysis, transition metal deposition controls nucleation of the carbon nanostructure. (E) edge-on high magnification SEM view of synthesized CNT and (F) extended SEM view of synthesized CNTs. E. Bottom: C2CNTs ($CO_2$ to CNTs) provide high conductivity and superior structural and conductive materials for a range of applications including lightweight jets, strengthened bridges, wind turbines, and electric vehicle body compnents and batteries. Cell illustration S. Licht.

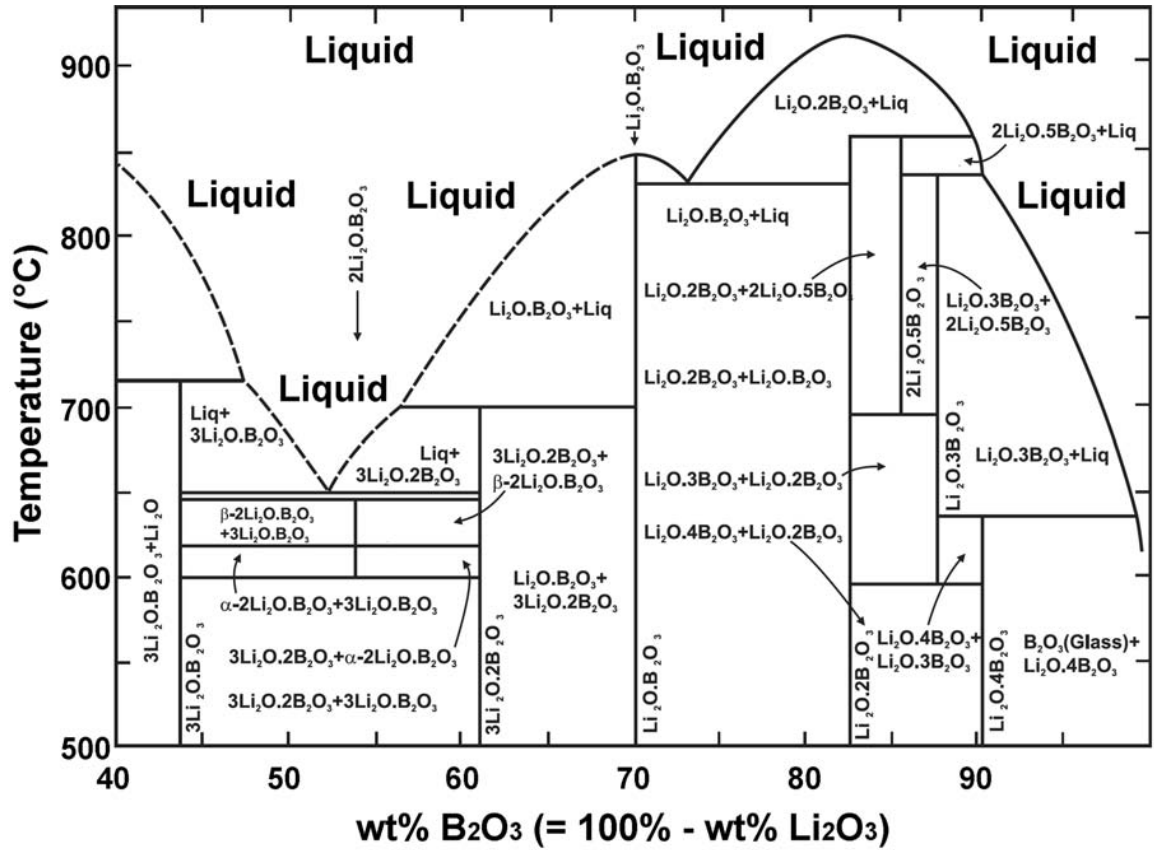

**Figure S2.** The liquid component of the binary Li₂O : B₂O₃ (wt%) system (From: ref. 26; modified from: Ferreira, E. B., Zanotto, E., Feller, S., Lodden, G., Banerjee, J., Edwards, T., Affatigat, M. Critical Analysis of Glass Stability Parameters and Application to Lithium Borate Glasses. J. Am. Ceram. Soc., **94**, 3833 (2011).).

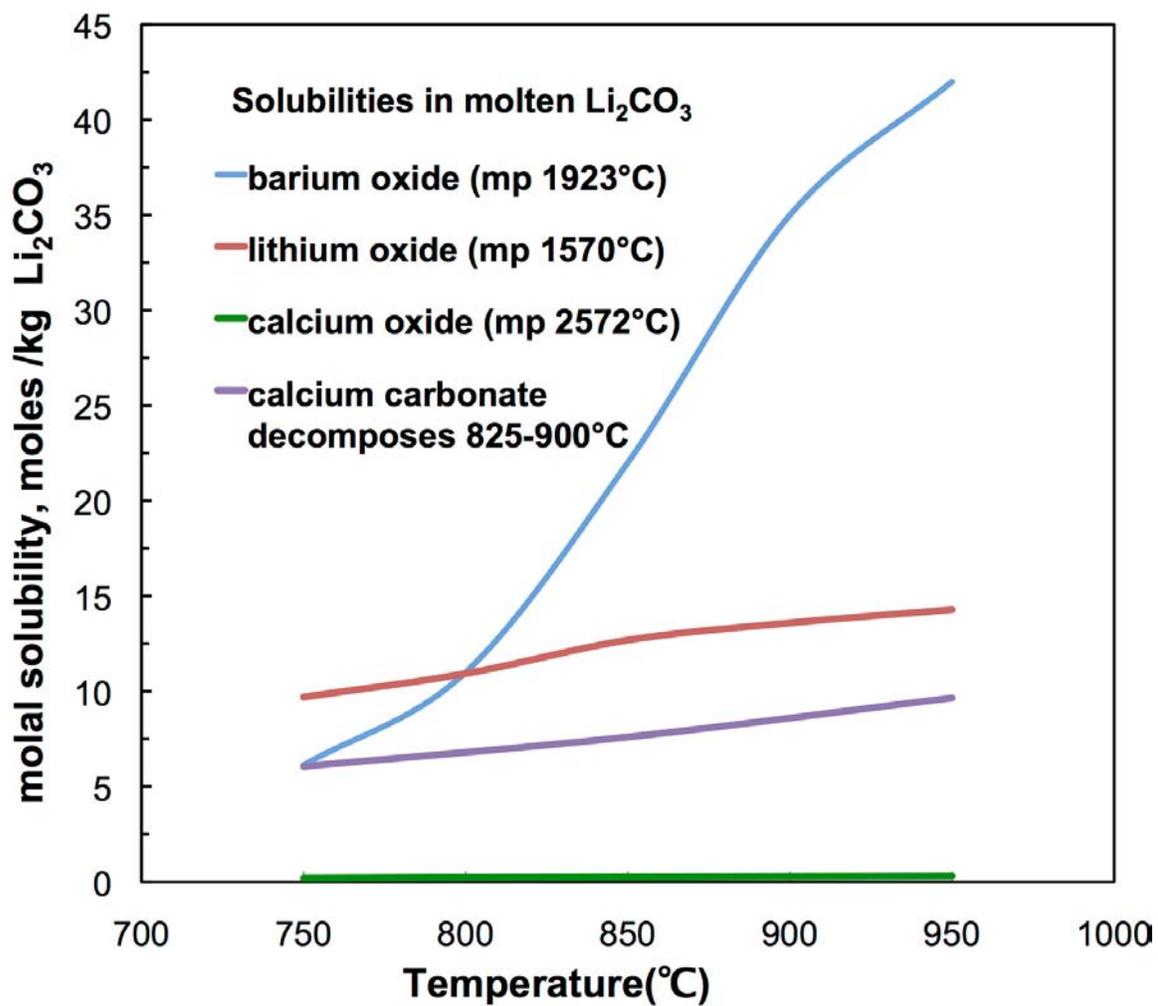

**Figure S3.** Solubilities in molten lithium carbonate of various oxides and calcium carobonate. From reference 27.